\documentclass[aip,jcp,twocolumn,superscriptaddress,groupedaddress]{revtex4}  
\usepackage{amsfonts}
\usepackage{amsmath}
\usepackage{amssymb}
\usepackage{version}
\usepackage{graphicx}%
\includeversion{New_connection}
\excludeversion{Old_connection}

\begin{document}
\title{Accurate Configurational and Kinetic Statistics in Discrete-Time Langevin Systems}
\author{Lucas Frese Gr{\o}nbech Jensen}
\affiliation{Department of Mathematics, University of California, Santa Barbara, CA 93106, USA}
\affiliation{Current Address: Department of Applied Mathematics and Computer Science, The Technical University of Denmark, 2800 Kgs.~Lyngby, Denmark}
\author{Niels Gr{\o}nbech-Jensen}
\affiliation{Department of Mechanical and Aerospace Engineering, University of California, Davis, CA 95616, USA}
\affiliation{Department of Mathematics, University of California, Davis, CA 95616, USA}
\begin{abstract}
We expand on the previously published Gr{\o}nbech-Jensen Farago (GJF) thermostat, which is a thermodynamically sound variation on the St{\o}rmer-Verlet algorithm for simulating discrete-time Langevin equations. The GJF method has been demonstrated to give robust and accurate configurational sampling of the phase space, and its applications to, e.g., Molecular Dynamics is well established. A new definition of the discrete-time velocity variable is proposed based on analytical calculations of the kinetic response of a harmonic oscillator subjected to friction and noise. The new companion velocity to the GJF method is demonstrated to yield correct and time-step-independent kinetic responses for, e.g., kinetic energy, its fluctuations, and Green-Kubo diffusion based on velocity autocorrelations. This observation allows for a new and convenient Leap-Frog algorithm, which efficiently and precisely represents statistical measures of both kinetic and configurational properties at any time step within the stability limit for the harmonic oscillator. We outline the simplicity of the algorithm and demonstrate its attractive time-step-independent features  for nonlinear and complex systems through applications to a one-dimensional nonlinear oscillator and three-dimensional Molecular Dynamics.
\end{abstract}
Submitted to Molecular Physics November 1, 2018\\ Accepted for publication January 7, 2019.
\maketitle
\section{Introduction}
\label{sec:intro}
Molecular Dynamics (MD) offers the possibility for simulating a set of $N$ interacting particles over a representative time period, and thereby producing trajectories that can be used to generate statistical and thermodynamic properties \cite{AllenTildesley,Frenkel,Rapaport}. This is accomplished by formulating a set of Newtonian equations of motion of the form
\begin{eqnarray}
m\ddot{r} & = & f \; ,\label{eq:NII}
\end{eqnarray}
where $m$ is the mass of a particle with coordinate $r$, which is subjected to the force $f$ that represents the particle's interaction with its surroundings. The discrete-time numerical approximation to the solution of Eq.~(\ref{eq:NII}) is most often acquired by methods rooted in the
St{\o}rmer-Verlet (SV) \cite{Stormer_1921,Verlet,Gear}
algorithm that evolves the coordinate $r(t_n)=r^n$ at time $t_n$ to the next time step, $dt$ later, at $t_{n+1}=t_n+dt$ when the coordinate is $r^{n+1}=r(t_{n+1})$. The second order (in $dt$) discrete-time equation is
\begin{eqnarray}
r^{n+1} & = & 2r^n-r^{n-1}+\frac{dt^2}{m}f^n \; ,\label{eq:SV}
\end{eqnarray}
where $f^n=f(r^n,t_n)$. This method is appealing for a number of reasons, including its simplicity, efficiency, stability properties, and time reversibility, which implies attractive conservation properties for the trajectory $r^n$ for closed systems ($f=f(r)$). A simple model system that directly illuminates the features of this algorithm is the harmonic oscillator, for which the SV method Eq.~(\ref{eq:SV}) can produce a discrete-time harmonic oscillator trajectory for as long as the reduced time step $dt\Omega_0<2$, where $\Omega_0$ is the natural frequency of the oscillator\cite{Venneri,Pastor_88}. While the SV method will reproduce a perfect harmonic oscillator, the discrete-time oscillation frequency $\Omega_V$ will be enhanced beyond the continuous-time frequency $\Omega_0$ (see, e.g., Ref.~\cite{GJF1} and references therein, or the Appendix in this paper).

Given the attractive properties of the SV discrete-time trajectory $r^n$, it is desirable to also approximate the accompanying velocity. This, however, introduces a new set of errors since discrete-time does not offer a proper definition of a differential, which is clearly necessary for evaluating the velocity. The two most obvious approximations are given by second order central differences. One of these \cite{AllenTildesley,Swope,Beeman} defines an {\it on-site} velocity $v^n$ by
\begin{eqnarray}
v^n & = & \frac{r^{n+1}-r^{n-1}}{2dt} \; , \label{eq:SV_v_basic}
\end{eqnarray}
which, combined with Eq.~(\ref{eq:SV}), can be expressed as the Velocity-explicit Verlet (VV) method 
\begin{eqnarray}
r^{n+1} & = & r^n+dt\,v^n+\frac{dt^2}{2m}f^n \label{eq:VV_r_basic}\\
v^{n+1} & = & v^n+\frac{dt^2}{2m}(f^n+f^{n+1}) \; . \label{eq:VV_v_basic}
\end{eqnarray}
The other approximation \cite{AllenTildesley,Buneman,Hockney} defines a {\it half-step} velocity $v^{n+\frac{1}{2}}$ by
\begin{eqnarray}
v^{n+\frac{1}{2}} & = & \frac{r^{n+1}-r^n}{dt}\; , \label{eq:lf_vel}
\end{eqnarray}
which, combined with Eq.~(\ref{eq:SV}), can be expressed as the Leap-Frog (LF) method
\begin{eqnarray}
v^{n+\frac{1}{2}} & = & v^{n-\frac{1}{2}}+\frac{dt}{m}f^n \label{eq:LF_v_basic}\\
r^{n+1} & = & r^n+dt\, v^{n+\frac{1}{2}} \; .\label{eq:LF_r_basic}
\end{eqnarray}
While the trajectories of Eqs.~(\ref{eq:SV}), (\ref{eq:VV_r_basic}), and (\ref{eq:LF_r_basic}) are identical, the associated velocity variables are different as can be seen from the following relationship
\begin{eqnarray}
v^n & = & \frac{1}{2}(v^{n-\frac{1}{2}}+v^{n+\frac{1}{2}}) \; . \label{eq:LF_v2_basic}
\end{eqnarray}
Of course, both velocity definitions limit the true velocity $v=\dot{r}$ for $dt\rightarrow0$.

Having highlighted the attractive features of the St{\o}rmer-Verlet method above, it is important to emphasize that the discrete-time velocity variables $v^n$ and $v^{n+\frac{1}{2}}$ are fundamentally inconsistent with $r^n$; i.e., the discrete-time velocities do not precisely represent the conjugated variable of the simulated trajectory. This has previously been indicated in various ways (see, e.g., Refs.~\cite{vgb_1982,Pastor_88,holian95,GJF1,GJF3}) through harmonic oscillator analysis. Thus, while $r^n$, $v^n$, and $v^{n+\frac{1}{2}}$ are all second order approximations to the true values $r(t_n)$ and $v(t_n)$, the mutual inconsistency between the coordinate and its velocity results in the discrete-time momentum not being the exact conjugated variable to the simulated trajectory. The harmonic oscillator illuminates this clearly, and the appendix of this paper further emphasizes this critical issue by including the LF half-step velocity into the analysis, indicating that properties extracted from Verlet methods will generally show inconsistencies between kinetic and configurational measures in the same simulation.

The SV, VV, and LF methods allow for simulations of the constant energy, or $(N,V,E)$, micro-canonical ensemble, whereas the constant temperature $(N,V,T)$ ensemble is often more relevant. The well-known governing equation for this type of dynamics is given by the Langevin equation \cite{Langevin}
\begin{eqnarray}
m\dot{v}+\alpha\dot{r} & = & f+\beta \; , \label{eq:Langevin}
\end{eqnarray}
where linear friction is represented by the constant $\alpha\ge0$, which is related to the thermal fluctuations $\beta$, which can be chosen to be represented by the Gaussian distribution \cite{Parisi}
\begin{eqnarray}
\langle\beta(t)\rangle & = & 0 \\
\langle\beta(t)\beta(t^\prime)\rangle & = & 2\alpha k_BT \delta(t-t^\prime) \; , 
\end{eqnarray}
where $k_B$ is Boltzmann's constant and $T$ is the thermodynamic temperature.

 Many methods for controlling the temperature of a simulated system (ÒthermostatsÓ) have been developed, and most of them fall into two major categories: Deterministic (e.g., Nos{\'e}-Hoover \cite{Nose,Hoover}) and stochastic (Langevin) thermostats \cite{SS,BBK,Pastor_88,Vanden,ML,Paquet}. The deterministic approach includes additional degrees of freedom, which act as an energy reservoir and thereby mimic a thermal heat bath. A requirement for such method is that the temperature of a simulated system can be reliably measured in order for the method to interact properly with the heat-bath. The stochastic approach is to directly simulate the Langevin equation Eq.~(\ref{eq:Langevin}), which does not include additional degrees of freedom, but instead interacts with a heat-bath through the fluctuation-dissipation balance for $\alpha>0$. Since discrete-time tends to distort the conjugated relationship between the positional coordinate and its corresponding momentum (see, e.g., Appendix in Ref.~\cite{GJF3} and the Appendix in this paper), a common problem for all methods is that kinetic and configurational measures of temperature disagree, which is a concern for both the integrity of a simulation and the extraction of self-consistent information, which may depend on configurational as well as kinetic sampling. It is therefore imperative to understand how to properly define a kinetic measure consistent with the statistics of the trajectory.

This work will consider the stochastic Langevin approach, which was pioneered by, e.g., Refs.~\cite{SS,BBK,Pastor_88,Gunsteren} by including discrete-time friction and noise into the Verlet framework. As clearly analyzed in Ref.~\cite{Pastor_88}, the inconsistencies between kinetic and configurational properties persist into the statistical averages obtained from the discrete-time thermostats. Consistent with the harmonic oscillator analysis of the algorithms, the configurational temperature is typically found to be higher than the measured kinetic temperature. One method was identified in Ref.~\cite{Pastor_88} to provide the correct time-step independent kinetic energy for a harmonic oscillator, but the inconsistency with the configurational properties made the authors reason that this was not desirable. Instead, their analysis revealed that while another method \cite{BBK} displayed significant time-step errors in energies, these errors were the same for kinetic and configurational measures. This consistency in error made the authors recommend using the stochastic BBK method \cite{BBK}. Since then, much work has been devoted to the development of methods that minimize the time-step error on statistical measures (see, e.g., Refs.~\cite{ML,Goga_2012,vdSpoel_2005,GJF1}). A major development of accomplishing correct sampling of configurational statistics for the harmonic oscillator was reported in 2012 \cite{ML}. This method, however, did not provide correct configurational diffusion in discrete time. We here focus specifically on the GJF method \cite{GJF1}, since that method has been analytically shown to provide exact configurational thermodynamic response for both flat and harmonic potentials for any time step within the Verlet stability criterion. The method has further been demonstrated to give extremely robust configurational sampling in complex molecular dynamics environments of both soft and hard materials \cite{GJF1,GJF2,GJF3,GJF4}, and it has been implemented into the MD suite LAMMPS \cite{Plimpton,LAMMPS-Manual} as an option for a stochastic thermostat. As expected, the kinetic measures of, e.g., kinetic energy (temperature) are, however, not reliable due to the time-discretization. We here seek to address this fundamental problem by redefining the discrete-time velocity variable such that the revised GJF method can provide reliable statistical information simultaneously for configurational as well as kinetic properties.

\section{Discrete-Time Langevin Dynamics}
\label{sec_lab_II}
We adopt the well-documented GJF method that was derived in Ref.~\cite{GJF1} and further analyzed for complex applications  Ref.~\cite{GJF2,GJF4}. The method reads
\begin{eqnarray}
r^{n+1} & = & r^n + b [dt\, v^n+\frac{dt^2}{2m}f^n+\frac{dt}{2m}\beta^{n+1}] \label{eq:gjf_r} \\
v^{n+1} & = & a\, v^n+\frac{dt}{2m}(af^n+f^{n+1})+\frac{b}{m}\beta^{n+1} \, , \label{eq:gjf_v}
\end{eqnarray}
where
\begin{eqnarray}
a & = & \frac{\displaystyle{1-\frac{\alpha dt}{2m}}}{\displaystyle{1+\frac{\alpha dt}{2m}}} \label{eq:a} \\
b & = & \frac{\displaystyle{1}}{\displaystyle{1+\frac{\alpha dt}{2m}}} \, . \label{eq:b}
\end{eqnarray}
The discrete-time noise is
\begin{eqnarray}
\beta^{n+1} & = & \int_{t_n}^{t_{n+1}}\beta(t^\prime)\,dt^\prime \, , \label{eq:discrete_beta}
\end{eqnarray}
which results in an uncorrelated Gaussian random number with zero mean and a variance given by the temperature and friction coefficient:
\begin{eqnarray}
\langle\beta^n\rangle & = & 0 \label{eq:noise_dis_ave} \\
\langle\beta^n\beta^l\rangle & = & 2\alpha k_BT dt \delta_{n,l} \, . \label{eq:noise_dis_std}
\end{eqnarray}
Notice that the special case $\alpha=0$ reduces the method to the VV form of the standard St{\o}rmer-Verlet algorithm, Eqs.~(\ref{eq:VV_r_basic}) and (\ref{eq:VV_v_basic}).
The derivation \cite{GJF1} of the GJF method was conducted by directly integrating the Langevin equation Eq.~(\ref{eq:Langevin}) over a time-step. By taking advantage of the normal distribution of the fluctuations and the exact integral of the friction term only as a function of spatial displacement, this approach maintains the fluctuation-dissipation relationship in discrete time. Thus, the known inaccuracies of velocity are avoided in the fluctuation-dissipation balance, and the configurational Boltzmann statistics remains accurate even for large time steps within the stability range (see, e.g., Eq.~(\ref{eq:gjf_epot}) below).
As outlined in Ref.~\cite{GJF2}, Eqs.~(\ref{eq:gjf_r}) and (\ref{eq:gjf_v}) can be combined to give the trajectory in the SV form, where the trajectory $r^n$ is separated from the associated velocity $v^n$ by linear transformation. The result is:
\begin{eqnarray}
r^{n+1} & = & 2br^n-ar^{n-1}+\frac{b\,dt^2}{m}f^n+\frac{b\,dt}{2m}(\beta^n+\beta^{n+1}) \nonumber \\\label{eq:gjf_sv}
\end{eqnarray}
where the GJF velocity Eq.~(\ref{eq:gjf_v}) can be expressed directly from the trajectory by \cite{note_on_typo}
\begin{eqnarray}
v^n & = & \frac{r^{n+1}-(1-a)r^n-ar^{n-1}}{2dt \, b}+\frac{1}{4m}(\beta^n-\beta^{n+1}) \; . \nonumber \label{eq:gjf_sv_v}\\
\end{eqnarray}
Thus, Eqs.~(\ref{eq:gjf_sv}) and (\ref{eq:gjf_sv_v}) can be recombined to yield Eqs.~(\ref{eq:gjf_r}) and (\ref{eq:gjf_v}), implying that the trajectory Eq.~(\ref{eq:gjf_sv}) is identical to what results from Eqs.~(\ref{eq:gjf_r}) and (\ref{eq:gjf_v}).

We will now give a new definition of a useful velocity parameter, based on a linear analysis of the GJF trajectory, Eq.~(\ref{eq:gjf_sv}).

\subsection{Linear Analysis, $f=-\kappa r$}
The introduction of the GJF algorithm \cite{GJF1} was rooted in the remarkable exact features for the linear Langevin equation (\ref{eq:Langevin}) for $f=-\kappa r$, $\kappa>0$, and $\alpha>0$:
\begin{eqnarray}
\langle E_p\rangle & = & \frac{1}{2}\kappa\langle(r^n)^2\rangle \; = \; \frac{1}{2}k_BT\label{eq:gjf_epot}\\
\langle E_k\rangle & = & \frac{1}{2}m\langle(v^n)^2\rangle \; = \; \frac{1}{2}k_BT\left(1-\frac{(\Omega_0dt)^2}{4}\right) \; , \label{eq:gjf_ekin}
\end{eqnarray}
with the additional observations that for $f=0$ ($\kappa=\Omega_0=0$) and $\alpha>0$, the method yields the exact Einstein diffusion $D=k_BT/\alpha$, and the correct kinetic energy given by Eq.~(\ref{eq:gjf_ekin}) for $\Omega_0=0$. These results are true for all $\Omega_0dt\le2$.

Equation (\ref{eq:gjf_ekin}) displays the features of the imperfect velocity given in Eq.~(\ref{eq:gjf_v}). Thus, while Eq.~(\ref{eq:gjf_epot}) is always an excellent measure of the configurational sampling, (\ref{eq:gjf_ekin}) is an increasingly poor representation of the kinetic temperature for increasing time-step $dt$. Of course, since the error factor in the measured kinetic energy is explicitly given, one is tempted to redefine the velocity in order to obtain the correct kinetic energy (temperature). However, that will only work for the specific harmonic mode with frequency $\Omega_0$ and a velocity correction of this kind can therefore not be successful for more complex or nonlinear problems.

Inspired by Ref.~\cite{Pastor_88}, where more than one velocity definition was applied to a given trajectory, we evaluate the kinetic energy of the GJF trajectory Eq.~(\ref{eq:gjf_sv}) based on the half-step velocity definition $v^{n+\frac{1}{2}}$ given in Eq.~(\ref{eq:lf_vel}):
\begin{eqnarray}
\langle E_k^{n+\frac{1}{2}}\rangle & = & \frac{1}{2}m\langle(v^{n+\frac{1}{2}})^2\rangle \; = \; \frac{1}{2}m\langle(\frac{r^{n+1}-r^n}{dt})^2\rangle \\
& = & \frac{m}{2dt^2}\left(\langle(r^{n+1})^2\rangle+\langle(r^{n})^2\rangle-2\langle r^{n+1}r^n\rangle\right)\nonumber \\
& = & \frac{m}{dt^2}\left(\langle(r^{n})^2\rangle-\langle r^{n+1}r^n\rangle\right)\, , \label{eq:ekin2}
\end{eqnarray}
where we have used $\langle(r^{n+1})^2\rangle=\langle(r^n)^2\rangle$.

For the harmonic oscillator, where $f=-\kappa r$, $\langle(r^n)^2\rangle$ is known for the GJF trajectory to be given by Eq.~(\ref{eq:gjf_epot}), $\langle r^{n+1}r^n\rangle$ can be found from Eq.~(\ref{eq:gjf_sv}) as follows:
Inserting $f=-\kappa r$ into Eq.~(\ref{eq:gjf_sv}) yields the autocorrelation
\begin{eqnarray}
\langle(r^{n+1})^2\rangle & = & 4b^2(1-\frac{(\Omega_0dt)^2}{2})^2 \langle(r^n)^2\rangle+a^2\langle(r^{n-1})^2\rangle \nonumber \\
& + & \left(\frac{b\,dt}{2m}\right)^2(\langle(\beta^n)^2\rangle+\langle(\beta^{n+1})^2\rangle) \nonumber \\
& + & 4b(1-\frac{(\Omega_0dt)^2}{2})\frac{b\,dt}{2m}\langle r^n\beta^n\rangle \nonumber \\
& - & 4ab(1-\frac{(\Omega_0dt)^2}{2})\langle r^{n}r^{n-1}\rangle \; . 
\end{eqnarray}
Since $\langle r^n\beta^n\rangle=\langle(\beta^n)^2\rangle bdt/2m$ (see Eq.~(\ref{eq:gjf_sv}) for $\langle r^{n+1}\beta^{n+1}\rangle$), we have from Eqs.~(\ref{eq:noise_dis_std}) and (\ref{eq:gjf_epot}) that
\begin{eqnarray}
\frac{k_BT}{\kappa} & = & \left[4b^2(1-\frac{(\Omega_0dt)^2}{2})^2+a^2\right]\frac{k_BT}{\kappa} \nonumber \\
& + & 2b^2(\Omega_0dt)^2\left[1+2b(1-\frac{(\Omega_0dt)^2}{2})\right]\frac{\alpha dt}{2m}\frac{k_BT}{\kappa}\nonumber\\
& - & 4ab(1-\frac{(\Omega_0dt)^2}{2})\langle r^{n}r^{n-1}\rangle \; , 
\end{eqnarray}
which, after some reduction, yields
\begin{eqnarray}
\langle r^nr^{n-1}\rangle & = & \langle r^nr^{n+1}\rangle \nonumber \\ & = & \frac{k_BT}{\kappa}\frac{4-2(\Omega_0dt)^2(1+b)+b(\Omega_0dt)^4}{4(1-\frac{(\Omega_0dt)^2}{2})} \; .\label{eq:cor_n_n+1}
\end{eqnarray}
Finally, inserting Eqs.~(\ref{eq:gjf_epot}) and (\ref{eq:cor_n_n+1}) into Eq.~(\ref{eq:ekin2}) gives
\begin{eqnarray}
\langle E_k^{n+\frac{1}{2}}\rangle & = & \frac{b}{2}k_BT \; . \label{eq:ekin2_final}
\end{eqnarray}
This is a useful result since it is independent of $\Omega_0 dt$. Thus, for {\it any} dynamical mode, we can for the GJF method propose the following re-assigned definition of discrete-time velocity:
\begin{eqnarray}
u^{n+\frac{1}{2}} & = & \frac{r^{n+1}-r^n}{\sqrt{b}\,dt} \; . \label{eq:gjf_vnew}
\end{eqnarray}
For the noisy harmonic oscillator, this 2GJ definition of velocity results in the average kinetic energy
\begin{eqnarray}
\langle E_k^{n+\frac{1}{2}}\rangle & = & \frac{1}{2}m\langle(u^{n+\frac{1}{2}})^2\rangle \; = \; \frac{1}{2}k_BT \; .  \label{eq:gjf_kin}
\end{eqnarray}
Thus, we now have self-consistency between time-step independent configurational and kinetic statistics for the noisy harmonic oscillator, and we have the correct statistical values of both potential and kinetic energies.

\subsection{Specific Verlet-type algorithms for using the 2GJ velocity}
The linear analysis above suggests that the Langevin equation with constant friction coefficient $\alpha$, and a general force of the kind $f=f(r)$, can be thermodynamically analyzed with some improvements by incorporating the re-assigned velocity $u^{n+\frac{1}{2}}$ into the GJF algorithm.
 
Direct applications of the accompanying velocity $u^{n+\frac{1}{2}}$ is straightforward for the GJF method in the SV form Eq.~(\ref{eq:gjf_sv}), since statistics of kinetic measures are easily conducted by $u^{n+\frac{1}{2}}$ from Eq.~(\ref{eq:gjf_vnew}). Similarly, the application of $u^{n+\frac{1}{2}}$ is straightforward for the GJF equations in the VV form Eqs.~(\ref{eq:gjf_r}) and (\ref{eq:gjf_v}), where the on-site GJF velocity $v^n$ is used for facilitating the calculation of $r^n$, and $u^{n+\frac{1}{2}}$ is used for calculating the kinetic properties.

Since the new measure of velocity is defined at the half-step, we can write an LF version of the GJF trajectory to directly take advantage of the 2GJ definition. Combining Eq.~(\ref{eq:gjf_vnew}) with the SV expression, Eq.~(\ref{eq:gjf_sv}), we obtain the GJF-2GJ algorithm
\begin{eqnarray}
u^{n+\frac{1}{2}} & = & au^{n-\frac{1}{2}}+\frac{\sqrt{b}\,dt}{m}f^n+\frac{\sqrt{b}}{2m}(\beta^n+\beta^{n+1})\label{eq:LGJ2_v}\\
r^{n+1} & = & r^n+\sqrt{b}\,dt\,u^{n+\frac{1}{2}} \; , \label{eq:LGJ2_r}
\end{eqnarray}
or, equivalently
\begin{eqnarray}
r^{n+1} & = & r^n+a\sqrt{b}\,dt\,u^{n-\frac{1}{2}}+\frac{b\,dt^2}{m}f^n+\frac{b\,dt}{2m}(\beta^n+\beta^{n+1}) \nonumber \\ \label{eq:LGJ2_rr}\\
u^{n+\frac{1}{2}} & = & \frac{r^{n+1}-r^n}{\sqrt{b}\,dt}\label{eq:LGJ2_vv}
\end{eqnarray}
where Eq.~(\ref{eq:LGJ2_vv}) is simply the 2GJ velocity definition from Eq.~(\ref{eq:gjf_vnew}).
Thus, one can express the method we are proposing here in many different convenient ways. However, the result is always the same; the hybrid between the GJF trajectory and the re-assigned 2GJ half-step velocity presented above. For $\alpha=0$, the method reduces to the standard LF form given in Eqs.~(\ref{eq:LF_v_basic}) and (\ref{eq:LF_r_basic}).

\subsection{Diffusion for $f=0$}

As mentioned above, the GJF trajectory has been shown to give the correct diffusion as calculated from the configurational Einstein definition of the diffusion coefficient $D$ (for $f=0$):
\begin{eqnarray}
D & = & \lim_{n\rightarrow\infty}\frac{\langle(r^{q+n}-r^q)^2\rangle_q}{2dt\,n} \; = \; \frac{k_BT}{\alpha}\label{eq:Einstein_D}
\end{eqnarray}
This is the true diffusion coefficient, since the measure considers the actual square displacement as a function of discrete time. The corresponding continuous-time kinetic Green-Kubo \cite{GreenKubo} expression for the same diffusion coefficient, calculated from
\begin{eqnarray}
D & = & \int_{0}^{\infty}\langle v(t_q+s)v(t_q)\rangle_q\,ds \; . \label{eq:Green-Kubo_cont}
\end{eqnarray}
where $v(t)$ is the continuous-time velocity.

Using the 2GJ velocity $u^{n+\frac{1}{2}}$ Eq.~(\ref{eq:LGJ2_v}) for $f=0$ we have the thermodynamically limiting form
\begin{eqnarray}
u^{q+\frac{1}{2}} & = & a^qu^\frac{1}{2}\nonumber \\
&+&\frac{\sqrt{b}}{2m}\left[a^{q-1}\beta^1+\beta^{q+1}+\sum_{k=0}^{q-2}a^k(a+1)\beta^{q-k}\right]\nonumber \\
&\rightarrow& \frac{\sqrt{b}}{2m}\left[\beta^{q+1}+\sum_{k=0}^{q-2}a^k(a+1)\beta^{q-k}\right]
\end{eqnarray}
for $|a|^q\rightarrow0$, which represents the thermodynamic limit (infinite volume) for $t\rightarrow\infty$, where all inertial information about the initial condition $u^{\frac{1}{2}}$ has been lost. The velocity auto-correlation for equilibrated initial conditions ($|a|^q\ll1$) is then
\begin{eqnarray}
\langle u^{q+\frac{1}{2}+n}\,u^{q+\frac{1}{2}}\rangle_q & = & \left\{\begin{array}{llc} \displaystyle\frac{k_BT}{m} & , & n=0\\
\\
\displaystyle\frac{k_BT}{m}ba^{n-1} & , & n>0 \end{array}\right. \label{eq:u_auto_c}
\end{eqnarray}

There are two discrete-time approximations when adapting the Green-Kubo expression (\ref{eq:Green-Kubo_cont}) to the discrete-time velocity correlation function. One is the accuracy of the velocity variable, which we have emphasized above. The other is the discretization of the $ds$ integral, since a non-zero $dt$-discretization allows the use of any Riemann sum for as long as that sum converges to the correct result when $dt\rightarrow0$. Thus, in discrete-time there are a multiple of different results for the Green-Kubo calculation of diffusion for each definition of the velocity. Choosing the right-Riemann sum for approximating the integral Eq.~(\ref{eq:Green-Kubo_cont}) we get
\begin{eqnarray}
D & = & \sum_{n=1}^\infty \frac{k_BT}{m}ba^{n-1} \, dt \; = \; \frac{k_BT}{m}\frac{b\, dt}{1-a} \; = \; \frac{k_BT}{\alpha} \; , \label{eq:Green_Kubo_disc}
\end{eqnarray}
which clearly shows the correct Green-Kubo diffusion coefficient for the 2GJ velocity definition for any time step $dt$. Notice that other approximations of the integral may give different results for nonzero $dt$. For example, the left-Riemann sum yields $D_{lR}=k_BT(\frac{1}{\alpha b}+\frac{dt}{m})$ and the trapezoidal sum gives $D_{tr}=k_BT/\alpha b$. While all possible Riemann approximations give the correct limit for $dt\rightarrow0$, only the right Riemann sum is independent of $dt$. We observe that if the GJF trajectory is paired with the traditional LF half-step definition of velocity, as given by Eq.~(\ref{eq:lf_vel}), the correct time-step independent result for the Green-Kubo expression for diffusion is obtained for the trapezoidal approximation to the integral Eq.~(\ref{eq:Green-Kubo_cont}). However, that choice of defining the velocity would yield the incorrect ($dt$ dependent) result Eq.~(\ref{eq:ekin2_final}) for the average kinetic energy in a harmonic potential. Thus, for the 2GJ velocity definition Eq.~(\ref{eq:gjf_vnew}), we define the right-Riemann sum to be the proper discrete-time approximation to the Green-Kubo integral, and we have then established consistency between kinetic and configurational measures, both for transport ($f=0$) and for the sampling of distributions and energies when $f$ represents a linear Hooke's force.

We note parenthetically that the case of $f=constant$ is entirely equivalent to the special case $f=0$ above, if one adds a constant velocity drift term $f/\alpha$ to the solution.

\section{Numerical Simulations}
\label{sec_lab_III}
The above analysis and results show that the proposed algorithm gives the correct configurational and kinetic statistical sampling of a noisy harmonic oscillator regardless of the applied time step. It also shows that configurational and kinetic measures are correct and consistent for any time step when the potential is flat. In order to verify the usefulness of the 2GJ velocity definition as a companion to the GJF trajectory in nonlinear and complex systems, we show the results of characteristic simulations both for a one-dimensional nonlinear oscillator and for representative Molecular Dynamics examples.
\subsection{One-dimensional nonlinear oscillator}
\label{one-dimensional}
We consider a one-dimensional oscillator with the deterministic force $f$ given by the potential energy surface $E_p(r)$
\begin{eqnarray}
f & = & -\frac{\partial E_p}{\partial r} \; , \label{eq:holonomic}
\end{eqnarray}
where $E_p$ is chosen to be a nonlinear convex potential
\begin{eqnarray}
\frac{E_p(r)}{E_0} & = & \left(\frac{r_0}{r_0-|r|}\right)^{6}\left[\left(\frac{r_0}{r_0-|r|}\right)^{6}-2\right]+1 \; , \label{eq:nonlin_pot}
\end{eqnarray}
for $|r|<r_0$. The characteristic energy and distance are denoted $E_0$ and $r_0$, respectively. We define the characteristic time scale from the inverse of the characteristic frequency $\omega_0=\sqrt{E_0/m}/r_0$, which yields the small amplitude oscillation frequency $\Omega_0=\sqrt{E^{\prime\prime}_p(0)/m}=\sqrt{60}\,\omega_0$, and the characteristic velocity $v_0=r_0\omega_0$.

Simulations are conducted for thermodynamic temperatures $k_BT=0.1E_0$ and $k_BT=0.5E_0$, and for friction coefficients $\alpha=1\,m\omega_0$ and $\alpha=5\,m\omega_0$. Statistical averages are acquired over 10$^{10}$ time steps. The statistical measures are as follows: Configurational and kinetic temperatures are defined as 
\begin{eqnarray}
T_c & = & \frac{1}{k_B}\frac{\langle(\frac{\partial E_p}{\partial r})^2\rangle}{\langle\frac{\partial^2E_p}{\partial r^2}\rangle} \label{eq:temp_con}\\
T_k & = & \frac{2}{k_B}\langle E_k\rangle \; , \label{eq:temp_kin}
\end{eqnarray}
where $T_c$ is the configurational temperature \cite{Hirschfelder,Rickayzen} and $T_k$ is the kinetic temperature from Eq.~(\ref{eq:gjf_kin}). We further evaluate the variances $\sigma^2_p$ and $\sigma_k^2$ of the potential and kinetic energies, respectively:
\begin{eqnarray}
\sigma_p^2 & = & \langle E_p^2\rangle-\langle E_p\rangle^2 \\
\sigma_k^2 & = & \langle E_k^2\rangle-\langle E_k\rangle^2 \; .
\end{eqnarray}

\begin{figure}[t]
\centering
\scalebox{0.5}{\centering \includegraphics[trim={2cm 2.5cm 2cm 3.0cm},clip]{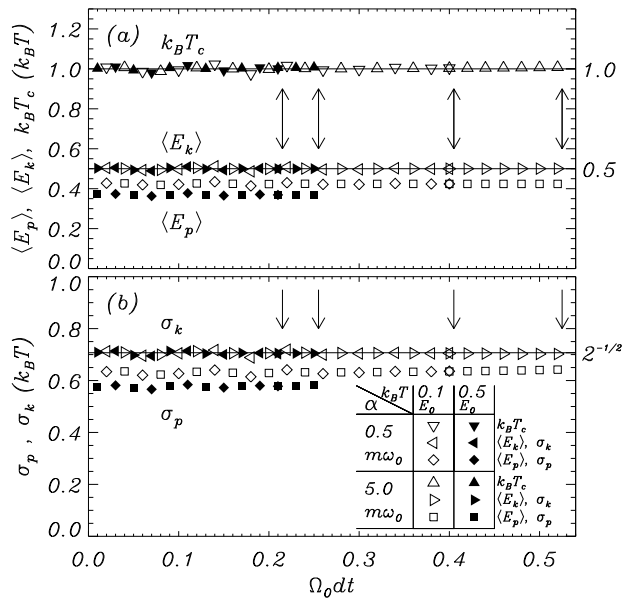}}
\caption{Statistical averages as a function of reduced time step $\Omega_0dt$ for the nonlinear oscillator described by the potential in Eq.~(\ref{eq:nonlin_pot}). Each marker is the result of 10$^{10}$ simulated time steps, and the parameter values represented by the markers are indicated in the inset table of (b). Vertical arrows indicate the approximate stability limits for $\Omega_0dt$. These are: $\Omega_0dt<0.215$ ($k_BT=0.5E_0$ and $\alpha=0.5m\omega_0$), $\Omega_0dt<0.255$ ($k_BT=0.5E_0$ and $\alpha=5m\omega_0$), $\Omega_0dt<0.405$ ($k_BT=0.1E_0$ and $\alpha=0.5m\omega_0$), $\Omega_0dt<0.525$ ($k_BT=0.1E_0$ and $\alpha=5m\omega_0$). (a): Statistical averages of configurational temperature $T_c$, kinetic and potential energies ($\langle E_k\rangle=\frac{1}{2}m\langle u^2\rangle$) and $\langle E_p(r)\rangle$.
All statistics of kinetic properties are calculated from the 2GJ velocity Eq.~(\ref{eq:gjf_vnew}).
Horizontal lines indicate the exact results $k_BT$ and $\frac{1}{2}k_BT$ for $k_BT_c$ and $\langle E_k\rangle$, respectively. (b): Standard deviations, $\sigma_k$ and $\sigma_p$, of the kinetic and potential energies, respectively. The known exact result $\frac{1}{\sqrt{2}}k_BT$ for $\sigma_k$ is given by the horizontal line.
}
\label{fig_1}
\end{figure}

Figure \ref{fig_1} shows the simulation results of these measures as a function of the reduced time step $\Omega_0dt$ for the four different parameter combinations of $T$ and $\alpha$ given above. The markers show the statistical data, while the horizontal lines indicate the correct theoretical results for $T_c=T$ and $\langle E_k\rangle=\frac{1}{2}k_BT_k=\frac{1}{2}k_BT$, and $\sigma_k=\frac{1}{\sqrt{2}}k_BT$. The inserted table indicates the symbols that represent the different parameter combinations and quantities. The arrows indicate the approximate stability limits for the four simulated parameter combinations. These limits are found by starting a simulation for small $\Omega_0dt$, then conducting 10$^{10}$ time steps before increasing $\Omega_0dt$ until the simulated particle displays instability by exceeding the bounds of the potential; i.e., the system is defined unstable when $|r|>r_0$. The stability limit results are: $\Omega_0dt<0.215$ ($k_BT=0.5E_0$ and $\alpha=0.5m\omega_0$), $\Omega_0dt<0.255$ ($k_BT=0.5E_0$ and $\alpha=5m\omega_0$), $\Omega_0dt<0.405$ ($k_BT=0.1E_0$ and $\alpha=0.5m\omega_0$), $\Omega_0dt<0.525$ ($k_BT=0.1E_0$ and $\alpha=5m\omega_0$). The different stability limits are due to the nonlinearity, which has a diverging curvature for increasing $|r|$. What is clear from the figure is that all statistical properties are nearly independent of the time step for as long as the stability limit has not been exceeded. Additionally, the known exact statistical values (indicated by horizontal lines and on the right legends) are very well matched by the simulations. None of the acquired data shows any deviating trends for increasing $\Omega_0dt$, which indicates that even the measures of potential energy $\langle E_p\rangle$ and its fluctuations $\sigma_p$ are correctly sampled, as has previously been demonstrated \cite{GJF2,GJF4}. We notice that simulations conducted for very low temperatures must result in a stability range $\Omega_0dt<2$, since the configurational sampling would effectively be happening in the harmonic limit.

\begin{figure}[t]
\centering
\scalebox{0.45}{\centering \includegraphics[trim={1.5cm 2.0cm 2cm 6.0cm},clip]{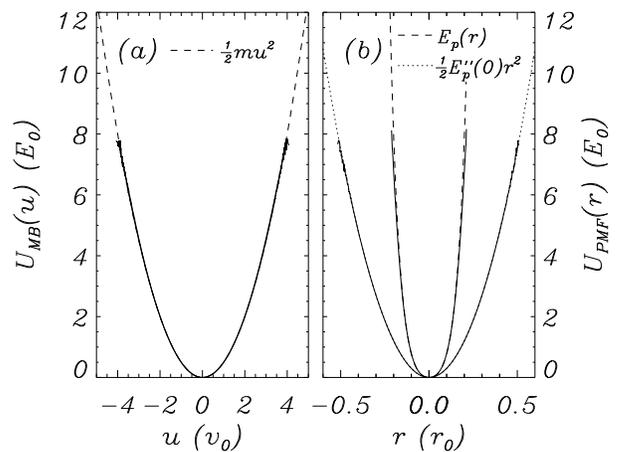}}
\caption{Effective potentials derived from simulated distributions of coordinate $r$ and velocity $u$ (Eq.~(\ref{eq:gjf_vnew})) for one dimensional oscillators. Solid curves are statistical results of Eqs.~(\ref{eq:MB}) and Eq.~(\ref{eq:PMF}) and dashed and dotted curves are the exact continuous-time results from the Maxwell-Boltzmann and Boltzmann distributions. Dashed/solid curves show results for the nonlinear oscillator of Eq.~(\ref{eq:nonlin_pot}) for $\Omega_0dt=0.52$ (see Figure \ref{fig_1}); dotted/solid curves show results of the harmonic oscillator with same small amplitude curvature for $\Omega_0dt=1.95$. Both selected time steps are very close to their respective stability limits. (a): Kinetic distributions. All results coincide on the figure. (b): Configurational distributions.
}
\label{fig_2}
\end{figure}

Encouraged by the robust statistical averages exemplified in Figure \ref{fig_1}, we explore if the sampling distributions of configurational coordinate $r$ and the 2GJ velocity $u$ match our expectations. We exemplify the results by showing data for $\alpha=5m\omega_0$ and $k_BT=0.5E_0$ for $\Omega_0dt=0.25$; i.e., the highest time step for which the system is stable. For this extreme time step we produce the configurational and kinetic density distributions $\rho_c(r)$ and $\rho_k(u)$ over the 10$^{10}$ simulated time steps. Ideally, these distributions relate to the potential and kinetic energies through the Maxwell and Maxwell-Boltzmann distributions such that
\begin{eqnarray}
\rho_c(r) & \propto & e^{-\frac{E_p(r)}{k_BT}} \label{eq:Boltzmann}\\
\rho_k(u) & \propto & e^{-\frac{E_k(u)}{k_BT}} \; , \label{eq:Maxwell-Boltzmann}
\end{eqnarray}
from where we can generate the effective potential and kinetic surfaces (potentials of mean force):
\begin{eqnarray}
U_{\rm PMF}(r) & = & -k_BT\ln\rho_c + C_c \label{eq:PMF}\\
U_{\rm MB}(u) & = & -k_BT\ln\rho_k + C_k \; . \label{eq:MB}
\end{eqnarray}
The constants $C_c$ and $C_k$ are determined such that, e.g., $E_{\rm PMF}(0)=E_p(0)$ and $U_{\rm MB}(0)=E_k(0)=0$. If the sampled discrete-time statistics is correct and consistent with the Boltzmann and Maxwell-Boltzmann distributions, we have the ideal comparisons: $U_{\rm PMF}(r)=E_p(r)$ and $U_{\rm MB}(u)=E_k(u)=\frac{1}{2}mu^2$.

Figure \ref{fig_2} shows the effective potentials $U_{\rm PMF}(r)$ and $U_{\rm MB}(u)$ for the parameters and procedures described above. Fig.~\ref{fig_2}a shows the simulation result $U_{\rm MB}(u)$ as a solid curve with the correct function $\frac{1}{2}mu^2$ shown as a dashed curve. It is obvious from this comparison that the 2GJ velocity is not only giving the correct, time-step independent statistical averages shown in Figure \ref{fig_1}, but also has the correct quadratic distribution even for this extreme value of the time step $\Omega_0dt$. Figure \ref{fig_2}b shows the comparable potentials of mean force $U_{\rm PMF}(r)$ as a solid curve with a comparison to the exact behavior $E_p(r)$ shown with a dashed curve. Again, this comparison is nearly flawless even for this very high time step. Included in the figure are the simulation results of a harmonic oscillator with the same parameters as for the nonlinear oscillator described above. We have chosen an oscillator with $\kappa=E_p^{\prime\prime}(0)$, and the results are shown for $\Omega_0dt=1.95$, which is very close to the stability limit for the harmonic oscillator. These results are shown as solid curves along with dotted curves that represent the exact results $\frac{1}{2}mu^2$ and $\frac{1}{2}\kappa r^2$. As expected from the derivation and analysis of the previous sections, we observe perfect agreement for the harmonic oscillator.

We conclude that the combination of the 2GJ velocity with the GJF trajectory yields very robust statistical results for both kinetic and configurational properties at any time step within the stability range, even for highly nonlinear systems.

\begin{figure}[t]
\centering
\scalebox{0.5}{\centering \includegraphics[trim={2cm 2.5cm 2cm 8.5cm},clip]{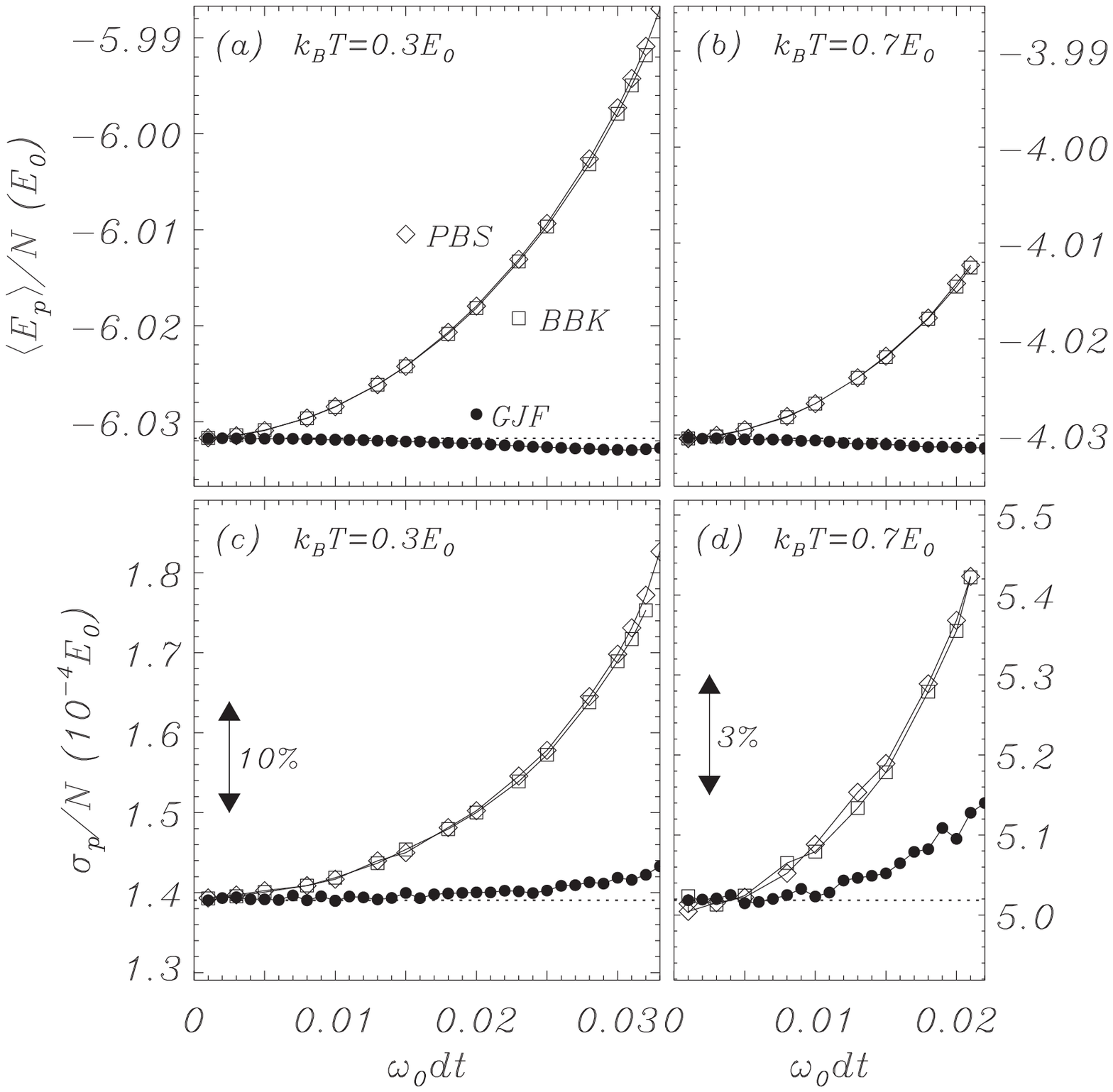}}
\caption{Statistical averages of potential energy $\langle E_p\rangle$ (a) and (b), and its standard deviation $\sigma_p$ (c) and (d) as a function of reduced time step $\omega_0dt$ for $\alpha=1\,m\omega_0$, sampled over $\omega_0\Delta t=2\times10^{5}$ units of time. $N=864$ particles are simulated with interaction potential Eq.~(\ref{eq:Eq_LJ_spline}) in a fixed cubic box with periodic boundary conditions. (a) and (c) show results for a crystalline FCC state at $k_BT=0.3E_0$ and volume $V=617.2558r_0^3$; (b) and (d) show results for a liquid state at $k_BT=0.7E_0$ and volume $V=824.9801r_0^3$. Results are shown for the GJF, BBK, and PBS trajectories. Horizontal dotted lines indicate the results for small $\omega_0dt$.
}
\label{fig_3}
\end{figure}

\begin{figure}[t]
\centering
\scalebox{0.5}{\centering \includegraphics[trim={2cm 2.5cm 2cm 8.5cm},clip]{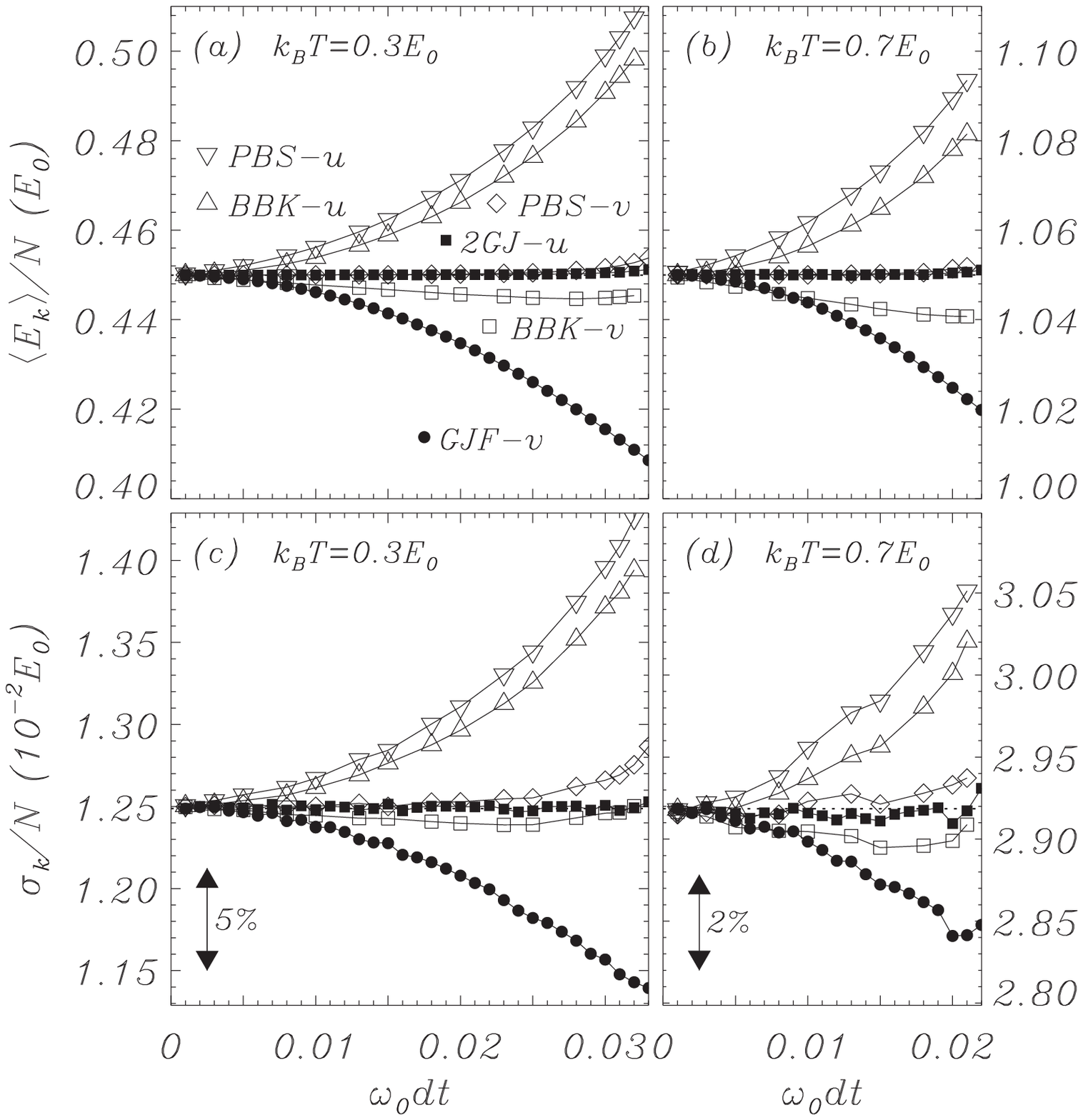}}
\caption{Statistical averages of kinetic energy $\langle E_k\rangle$, (a) and (b), and its standard deviation $\sigma_k$, (c) and (d), as a function of reduced time step $\omega_0dt$ for $\alpha=1\,m\omega_0$, sampled over $\omega_0\Delta t=2\times10^{5}$ units of time. $N=864$ particles are simulated with interaction potential Eq.~(\ref{eq:Eq_LJ_spline}) in a fixed cubic box with periodic boundary conditions. (a) and (c) show results for a crystalline FCC state at $k_BT=0.3E_0$ and volume $V=617.2558r_0^3$; (b) and (d) show results for a liquid state at $k_BT=0.7E_0$ and volume $V=824.9801r_0^3$. The shown axes with $\omega_0dt$ cover the ranges of stability.  Results are calculated from the GJF, BBK, and PBS trajectories for both on-site ($v$) and half-step ($u$) velocities. Horizontal dotted lines indicate the results for small $\omega_0dt$. Markers $\bullet$ represent GJF simulations with the GJF velocity given by Eq.~(\ref{eq:gjf_v}), labeled $GJF\!-\!v$. Markers $\blacksquare$ represent the results of the GJF trajectory with the 2GJ velocity Eq.~(\ref{eq:gjf_vnew}), labeled $2GJ\!-\!u$.
}
\label{fig_4}
\end{figure}

\begin{figure}[t]
\centering
\scalebox{0.5}{\centering \includegraphics[trim={2.0cm 2.5cm 2.cm 8.5cm},clip]{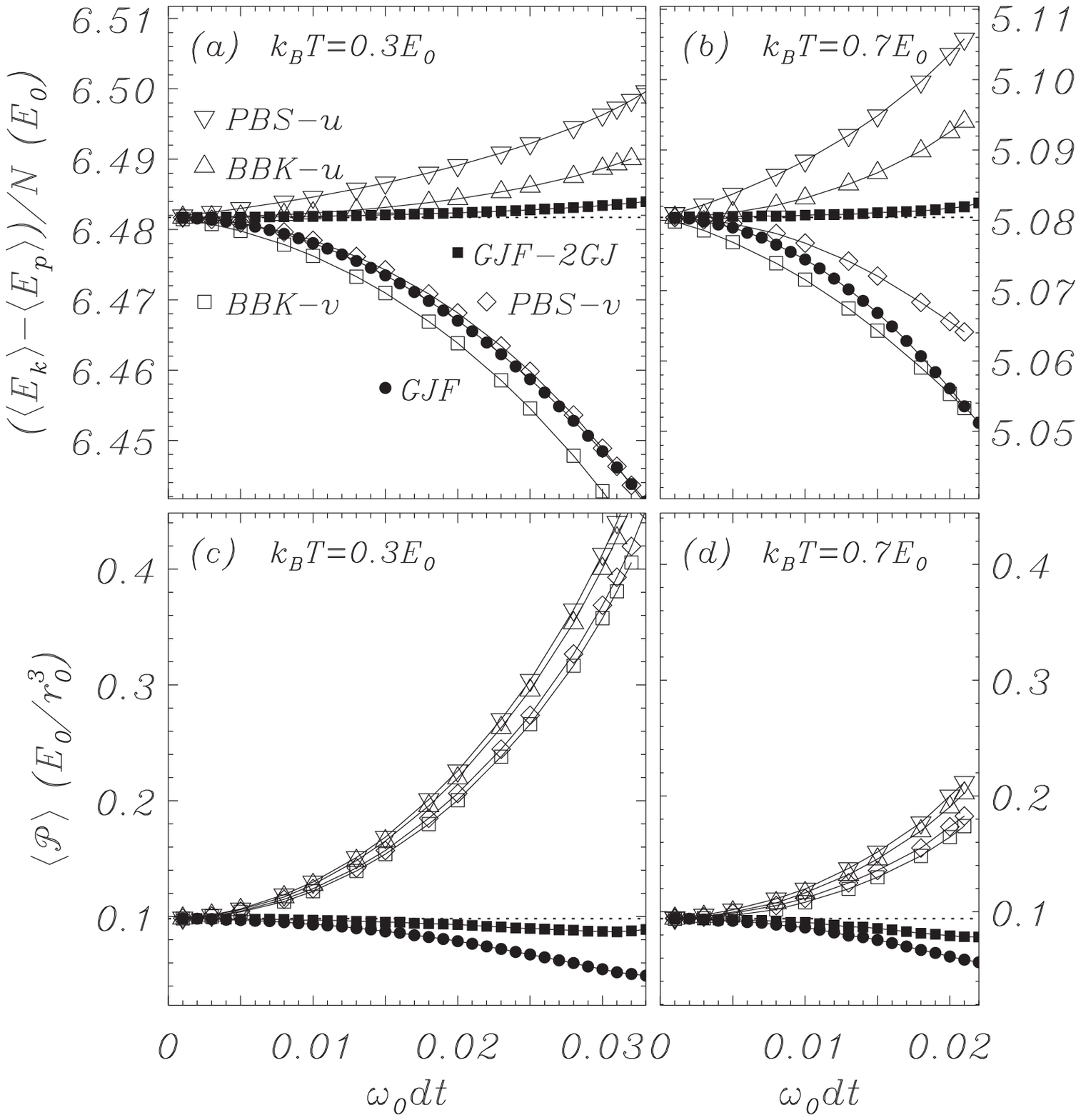}}
\caption{Statistical averages of combinations of kinetic and configurational measures as a function of reduced time step $\omega_0dt$ for $\alpha=1\,m\omega_0$, sampled over $\omega_0\Delta t=2\times10^{5}$ units of time. $N=864$ particles are simulated with interaction potential Eq.~(\ref{eq:Eq_LJ_spline}) in a fixed cubic box with periodic boundary conditions. $\langle E_k\rangle-\langle E_p\rangle$ (a) and (b), and system pressure ${\cal P}$, (c) and (d), calculated from Eq.~(\ref{eq:Press}). (a) and (c) show results for a crystalline FCC state at $k_BT=0.3E_0$ and volume $V=617.2558r_0^3$; (b) and (d) show results for a liquid state at $k_BT=0.7E_0$ and volume $V=824.9801r_0^3$. The shown axes with $\omega_0dt$ cover the ranges of stability.  Results are calculated from the GJF, BBK, and PBS trajectories for both on-site ($v$) and half-step ($u$) velocities. Horizontal dotted lines indicate the results for small $\omega_0dt$. Markers $\bullet$ represent GJF simulations with the GJF velocity given by Eq.~(\ref{eq:gjf_v}), labeled $GJF$. Markers $\blacksquare$ represent the results of the GJF trajectory with the 2GJ velocity Eq.~(\ref{eq:gjf_vnew}), labeled $GJF\!-\!2GJ$.
}
\label{fig_5}
\end{figure}

\subsection{Molecular Dynamics}
We wish to further demonstrate the proposed algorithm for use in Molecular Dynamics by showing characteristic simulation results for both solid and liquid states of a simple, one-component system in the $(N,V,T)$ ensemble. We adopt the splined, short range Lennard-Jones potential described in Ref.~\cite{GJF3}:
\begin{eqnarray}
\frac{E_p(|r|)}{E_0} & = &
\left\{\begin{array}{lcl} \displaystyle{\left(\frac{|r|}{r_0}\right)^{-12}-2\left(\frac{|r|}{r_0}\right)^{-6}} & , & 0<|r|\le r_s
\\ \displaystyle{\frac{a_4}{E_0}\left(|r|-r_c\right)^4+\frac{a_8}{E_0}\left(|r|-r_c\right)^8} & , & r_s<|r|<r_c \\ \displaystyle{0} & , & r_c \le |r|
\end{array}\right. \label{eq:Eq_LJ_spline}
\end{eqnarray}
where $r$ is a three dimensional coordinate between any two particles. The parameters are given by
\begin{eqnarray}
\frac{r_s}{r_0} & = & \left(\frac{13}{7}\right)^{1/6} \; \approx \; 1.108683 \\
\frac{r_c}{r_0} & = & \frac{r_s}{r_0}-\frac{32E_p(r_s)}{11E_p^\prime(r_s)r_0} \; \approx \; 1.959794\\
a_4 & = & \frac{8E_p(r_s)+(r_c-r_s)E_p^\prime(r_s)}{4(r_c-r_s)^4} \\
a_8 & = & -\frac{4E_p(r_s)+(r_c-r_s)E_p^\prime(r_s)}{4(r_c-r_s)^8} .
\end{eqnarray}
The potential $E_p(r)$ has minimum $-E_0$ at $|r|=r_0$, and is smoothly splined
between the inflection point of the Lennard-Jones potential and zero with continuity through the second derivative
at $|r|=r_s$, and continuity through the third derivative at $|r|=r_c$. We simulate $N=864$ particles of identical mass $m$ in a cubic box of side-length $L$, and with periodic boundary conditions. Initial conditions are chosen to be the Face Centered Cubic (FCC) structure, which is allowed to evolve for a long transient time before acquiring statistical data in order to reach a thermodynamically representative ensemble for a given applied temperature and a reduced time step $\omega_0dt$, where $\omega_0^2mr_0^2=E_0$.

We conduct simulations for two characteristically different temperatures, representing a crystalline solid at a set volume of $V=L^3=(8.51442r_0)^3$ for $k_BT=0.3E_0$, and a liquid at a set volume of $V=L^3=(9.378812r_0)^3$ for $k_BT=0.7E_0$. Each simulation makes statistical averages over a time span of $\omega_0\Delta t=2\times10^5$.

\begin{figure}[t]
\centering
\scalebox{0.5}{\centering \includegraphics[trim={2cm  2.5cm 2cm 8.5cm},clip]{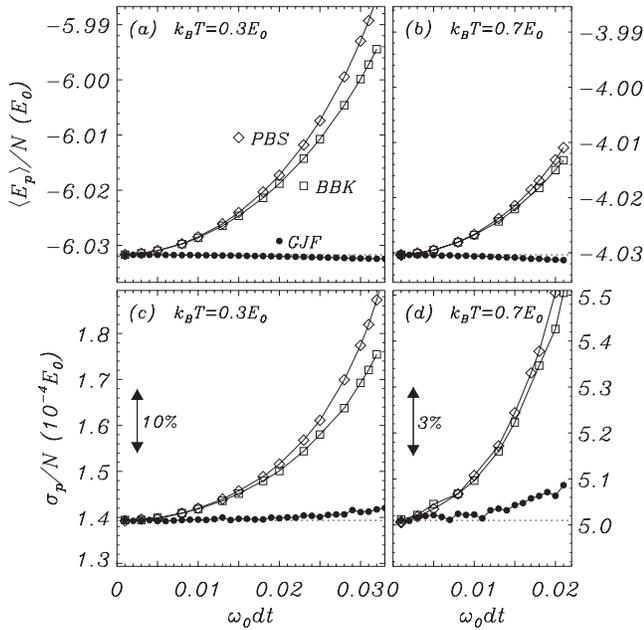}}
\caption{Statistical averages of potential energy $\langle E_p\rangle$, (a) and (b), and its standard deviation $\sigma_p$, (c) and (d), as a function of reduced time step $\omega_0dt$ for $\alpha=10\,m\omega_0$, sampled over $\omega_0\Delta t=2\times10^{5}$ units of time. $N=864$ particles are simulated with interaction potential Eq.~(\ref{eq:Eq_LJ_spline}) in a fixed cubic box with periodic boundary conditions. (a) and (c) show results for a crystalline FCC state at $k_BT=0.3E_0$ and volume $V=617.2558r_0^3$; (b) and (d) show results for a liquid state at $k_BT=0.7E_0$ and volume $V=824.9801r_0^3$. The shown axes with $\omega_0dt$ cover the ranges of stability.  Results are shown for the GJF, BBK, and PBS trajectories. Horizontal dotted lines indicate the results for small $\omega_0dt$.
}
\label{fig_6}
\end{figure}

\begin{figure}[t]
\centering
\scalebox{0.5}{\centering \includegraphics[trim={2cm 2.5cm 2cm 8.5cm},clip]{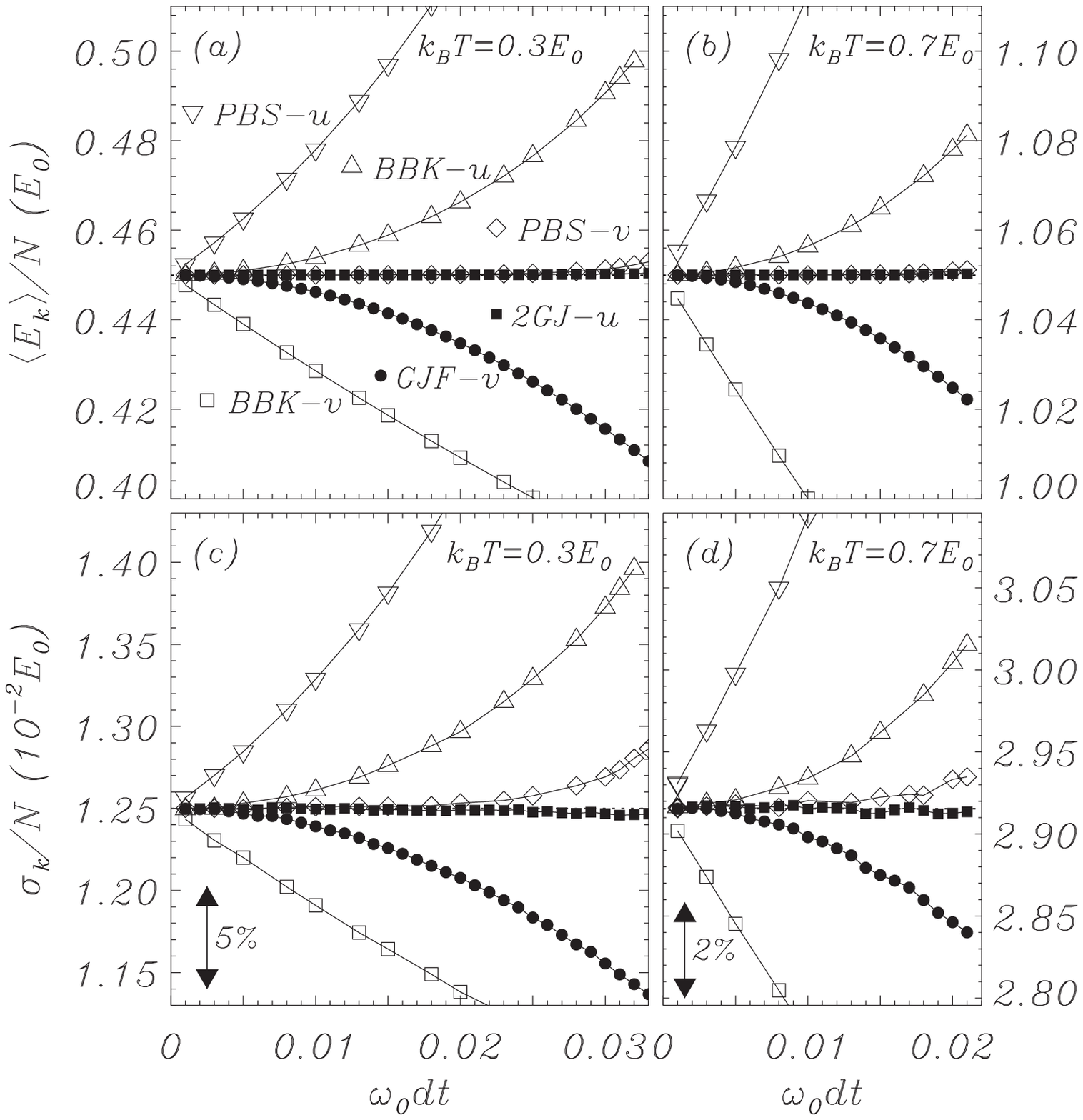}}
\caption{Statistical averages of kinetic energy $\langle E_k\rangle$, (a) and (b), and its standard deviation $\sigma_k$, (c) and (d), as a function of reduced time step $\omega_0dt$ for $\alpha=10\,m\omega_0$, sampled over $\omega_0\Delta t=2\times10^{5}$ units of time. $N=864$ particles are simulated with interaction potential Eq.~(\ref{eq:Eq_LJ_spline}) in a fixed cubic box with periodic boundary conditions. (a) and (c) show results for a crystalline FCC state at $k_BT=0.3E_0$ and volume $V=617.2558r_0^3$; (b) and (d) show results for a liquid state at $k_BT=0.7E_0$ and volume $V=824.9801r_0^3$. The shown axes with $\omega_0dt$ cover the ranges of stability.  Results are calculated from the GJF, BBK, and PBS trajectories for both on-site ($v$) and half-step ($u$) velocities. Horizontal dotted lines indicate the results for small $\omega_0dt$. Markers $\bullet$ represent GJF simulations with the GJF velocity given by Eq.~(\ref{eq:gjf_v}), labeled $GJF\!-\!v$. Markers $\blacksquare$ represent the results of the GJF trajectory with the 2GJ velocity Eq.~(\ref{eq:gjf_vnew}), labeled $2GJ\!-\!u$.
}
\label{fig_7}
\end{figure}

\begin{figure}[t]
\centering
\scalebox{0.5}{\centering \includegraphics[trim={2cm  2.5cm 2cm 8.5cm},clip]{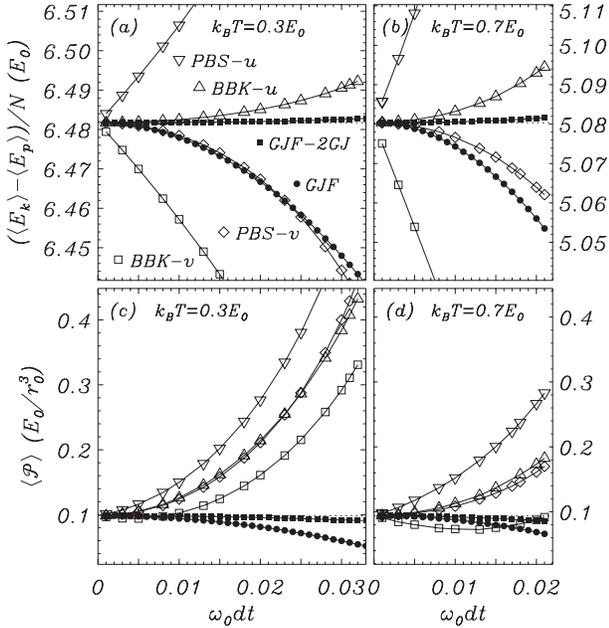}}
\caption{Statistical averages of combinations of kinetic and configurational measures as a function of reduced time step $\omega_0dt$ for $\alpha=10\,m\omega_0$, sampled over $\omega_0\Delta t=2\times10^{5}$ units of time. $N=864$ particles are simulated with interaction potential Eq.~(\ref{eq:Eq_LJ_spline}) in a fixed cubic box with periodic boundary conditions. $\langle E_k\rangle-\langle E_p\rangle$, (a) and (b), and system pressure ${\cal P}$, (c) and (d), calculated from Eq.~(\ref{eq:Press}). (a) and (c) show results for a crystalline FCC state at $k_BT=0.3E_0$ and volume $V=617.2558r_0^3$; (b) and (d) show results for a liquid state at $k_BT=0.7E_0$ and volume $V=824.9801r_0^3$. The shown axes with $\omega_0dt$ cover the ranges of stability. Results are calculated from the GJF, BBK, and PBS trajectories for both on-site ($v$) and half-step ($u$) velocities. Horizontal dotted lines indicate the results for small $\omega_0dt$. Markers $\bullet$ represent GJF simulations with the GJF velocity given by Eq.~(\ref{eq:gjf_v}), labeled $GJF$. Markers $\blacksquare$ represent the results of the GJF trajectory with the 2GJ velocity Eq.~(\ref{eq:gjf_vnew}), labeled $GJF\!-\!2GJ$.
}
\label{fig_8}
\end{figure}

The GJF and GJF-2GJ results shown in Figures \ref{fig_3}-\ref{fig_8} are labeled as follows: For purely configurational measures, shown in Figs.~\ref{fig_3} and \ref{fig_6}, the results are labeled {\it GJF}, since the GJF and GJF-2GJ methods differ only in the velocity. For purely kinetic measures, shown in Figs.~\ref{fig_4} and \ref{fig_7}, GJF results using the on-site GJF velocity in Eqs.~(\ref{eq:gjf_v}) are labeled {\it GJF-v}, while GJF results for the half-step 2GJ velocity of Eq.~(\ref{eq:gjf_vnew}) are labeled {\it GJF-u}. Finally, for composite measures of kinetic and configurational statistics, shown in Figs.~\ref{fig_5} and \ref{fig_8}, results using the original GJF method Eqs.~(\ref{eq:gjf_r}) and (\ref{eq:gjf_v}) are labeled $GJF$, while results using the GJF trajectory with the 2GJ velocity (e.g., Eqs.~(\ref{eq:LGJ2_v}) and (\ref{eq:LGJ2_r})) are labeled {\it GJF-2GJ}.

In addition to conducting MD simulations using the GJF and GJF-2GJ methods that we propose in this paper, we have chosen to include a select few of well studied and well-used traditional methods \cite{BBK,Loncharich_2004,Pastor_88} that exemplify the inconsistencies between configurational and kinetic properties in discrete time. There are many, more recent methods, which notably includes the key advance shown in Ref.~\cite{ML}, where the proper configurational Boltzmann distribution was first obtained. Given that the GJF method includes this essential feature, as well as other advances, we let the GJF trajectory represent this class of attractive methods. Our main objective here is to show that the GJF-2GJ combination gives very attractive results for MD in {\it both} configurational and kinetic sampling. 

Apart from the GJF and GJF-2GJ results of this paper, we will thus compare to four other sets of results. These are generated by the well-characterized methods given in Ref.~\cite{Pastor_88}: Equation (2.20) in Ref.~\cite{Pastor_88} with the on-site velocity Eq.~(\ref{eq:SV_v_basic}) (BBK-v); Equation (2.20) in Ref.~\cite{Pastor_88} with the half-step velocity Eq.~(\ref{eq:lf_vel}) (BBK-u); Equation (2.21) in Ref.~\cite{Pastor_88} with the on-site velocity Eq.~(\ref{eq:SV_v_basic}) (PBS-v); and Equation (2.21) in Ref.~\cite{Pastor_88} with the half-step velocity Eq.~(\ref{eq:lf_vel}) (PBS-u).

Figures~\ref{fig_3}-\ref{fig_5} show results for $\alpha=1\,m\omega_0$ as a function of the reduced time step $\omega_0dt$ for the entire ranges of stability. Figure \ref{fig_3}a confirms the attractive configurational sampling properties of the GJF method by displaying only very modest deviation from time-step independent behavior in average potential energies and their fluctuations compared to the traditional BBK and PBS methods. As expected, the liquid state (left panels (b) and (d)) at $k_BT=0.7E_0$ exhibits the most deviation (about 2-3\% near the stability range) since this state is a disordered ensemble with relatively slow diffusion, which makes the statistical sampling more challenging than for a crystalline state (right panels (a) and (c)) at $k_BT=0.3E_0$.

Figure \ref{fig_4} displays the attractive kinetic sampling properties of the GJF with the 2GJ velocity ($\blacksquare$). The time-step independence of the kinetic measures is striking for both kinetic energy averages and their fluctuations, even in the liquid state at $k_BT=0.7E_0$. We notice that the PBS-v and, to a lesser degree, the BBK-v results also display very good time-step independence. However, the comparable configurational properties from Figure \ref{fig_3}a reveals the internal inconsistencies of those methods. As pointed out in Ref.~\cite{Pastor_88}, the better use of BBK and PBS is to use BBK-u and PBS-u since these deviations in kinetic properties are somewhat consistent with the deviations in the BBK and PBS configurational measures.

Figure \ref{fig_5} investigates the relationships between the kinetic and configurational statistics with (a) and (b) exploring the difference $\langle E_k\rangle-\langle E_p\rangle$, and (c) and (d) showing the system pressure ${\cal P}$ calculated from
\begin{eqnarray}
{\cal P} & = & \frac{1}{3V}\left\langle\sum_{i=1}^Nf_i\cdot r_i\right\rangle+\frac{Nk_BT_k}{V} \; , \label{eq:Press}
\end{eqnarray}
where $f_i$ is the total force on the particle with coordinate $r_i$, and where $T_k$ is calculated from Eq.~(\ref{eq:temp_kin}).  Again, we observe that the GJF-2GJ combination ($\blacksquare$) is superior in its behavior as the time step is increased, with only minor deviations in both statistical energy difference and the pressure. We observe the GJF result ($\bullet$) deviating significantly for the energy differences, due to the known depression of the velocity magnitude, and this translates into the visible depression of the calculated pressure, although the GJF pressure is still better than the ones calculated from the BBK and PBS methods.

Figures \ref{fig_6}-\ref{fig_8} show the same kinds of results as in Figures \ref{fig_3}-\ref{fig_5}, but for a friction coefficient of $\alpha=10\,m\omega_0$. The overall observation from these figures is, as expected, much the same for GJF and GJF-2GJ, since these methods are shown in the earlier sections of this paper to have deviations independent of the friction coefficient. In fact, it is clear that the higher friction allows the system to be in more tight contact with the heat-bath, which results in better statistical averages. Only the fluctuation of the potential energy of the liquid state observed in Figure \ref{fig_6}d shows any appreciable deviation (2-3\%). The kinetic measures using the 2GJ velocity are as independent of the time step as one can expect. In contrast, we observe that the BBK and PBS results generally show significant deviations as the time step is increased. The expected exception are the PBS-v kinetic results seen in Figure \ref{fig_7}, but these very good kinetic properties are not matched by reliable configurational properties, as observed in Figure \ref{fig_6}.

\section{Discussion}
\label{sec_lab_IV}

We have presented, analyzed, and tested a new method for modeling Langevin equations in discrete time. The method is based on the trajectory of the GJF thermostat, which has been shown to possess exact configurational sampling properties in linear systems for any choice of time step within the stability limit. We have here devised a definition for a companion half-step velocity, which possesses equally good statistical sampling properties for kinetic measures. The combination of the two parts, the GJF trajectory and the 2GJ velocity, can be utilized in any of the formulated GJF forms, SV, VV, or LF. However, we have formulated a specific LF formulation that directly incorporates the 2GJ velocity into the GJF trajectory calculation, such that this new GJF-2GJ method conveniently displays all the attractive features of both.

Apart from the exact, time-step-independent results obtained for both configurational and kinetic sampling for linear systems, the method has been tested for a nonlinear oscillator subjected to friction and noise, and the results have confirmed near-perfect statistical behavior for as long as the trajectory is stable. Similarly, we have tested the method for MD $(N,V,T)$ ensembles in both crystalline and liquid states for different friction coefficients. Again, the promising features of the method are confirmed for both configurational and kinetic properties as well as for combinations of the two, as exemplified in Figs.~\ref{fig_5} and \ref{fig_8} by calculating differences between kinetic and potential energies as well as the pressure as a function of reduced time step with near perfect time-step independence of the results.

Given the simplicity of the method, along with the very familiar St{\o}rmer-Verlet form of the expressions, we submit that this is a very convenient method for ensuring the best possible statistical results from Langevin and MD systems. Linear analysis of other methods \cite{BBK,Skeel_2003,eb_1980,vgb_1982,Skeel_2002,rc_2003,Melchionna_2007,Bussi_2007,thalmann_2007,Mishra_1996,ML} have demonstrated good second and third order accuracy in the obtained thermodynamic quantities, but we are not aware of any other method that has been demonstrated to give the exact, time-step independent thermodynamic response for both position and velocity. Thus, the method allows for more accurate and more efficient acquisition of reliable statistical data at no additional computational cost compared to any other method we are aware of. The GJF integrator is currently available within the LAMMPS simulation package. Including the 2GJ velocity in simulations seems straightforward.

\appendix
\label{appndx}
\section{The SV Method Applied to The Deterministic Harmonic Oscillator}
The harmonic oscillator is given by Eq.~(\ref{eq:NII}) with $f=-\kappa r$, where $\kappa>0$ is the spring constant, and the natural frequency of the oscillator is therefore $\Omega_0=\sqrt{\kappa/m}$. The discrete time (complex) solution to, e.g., Eq.~(\ref{eq:SV}) yields \cite{GJF1}
\begin{eqnarray}
r^n & = & e^{\pm i\Omega_V dt \, n} \label{eq:harmonic_sol}
\end{eqnarray}
where the discrete-time frequency $\Omega_V>\Omega_0$ is given by
\begin{eqnarray}
\cos{\Omega_V dt} & = & 1-\frac{(\Omega_0dt)^2}{2} \label{eq:cosWV}\\
\sin{\Omega_V dt} & = & \Omega_0dt\sqrt{1-\frac{(\Omega_0dt)^2}{4}} \label{eq:sinWV}
\end{eqnarray}
Thus, it is obvious that for as long as $\Omega_0dt\le2$ (the stability limit) the trajectory $r^n$ is a perfect harmonic oscillator with a frequency $\Omega_V$ that quadratically approximates $\Omega_0$ for small $\Omega_0dt$ and becomes $\Omega_V\rightarrow\frac{\pi}{2}\Omega_0$ for $\Omega_0 dt\rightarrow2$.

Following the appendix of Ref.~\cite{GJF3}, inserting the harmonic solution Eq.~(\ref{eq:harmonic_sol}) into the on-site approximation to the velocity Eq.~(\ref{eq:SV_v_basic}) yields
\begin{eqnarray}
v^n & = & \pm i\Omega_V r^n \, \frac{\sin{\Omega_V dt}}{\Omega_V dt}  
\label{eq:harmonic_vn}
\end{eqnarray}
Similarly, inserting Eq.~(\ref{eq:harmonic_sol}) into the half-step approximation Eq.~(\ref{eq:lf_vel}) to the velocity yields
\begin{eqnarray}
v^{n+\frac{1}{2}} & = & \pm i\Omega_V r^{n+\frac{1}{2}} \, \frac{\sin{\frac{\Omega_V dt}{2}}}{\frac{\Omega_V dt}{2}} \label{eq:harmonic_vnh}
\end{eqnarray}
A consistent velocity should read $v^m=\pm i\Omega_Vr^m$, which means that both the VV velocity and the half-step LF velocity are inconsistent with the trajectory. It is further clear that both velocity definitions yield values that are depressed in magnitude for the harmonic oscillator (convex potential), and we can see that the half-step LF velocity is, as expected, considerably more accurate than the on-site VV velocity. The latter observation is particularly important if one wishes to calculate, e.g., an average kinetic energy (for assessing, e.g., a thermodynamic temperature).

Implied by this short review of the Verlet method applied to the harmonic oscillator is that the discrete-time velocity is {\it not} precisely the associated velocity of the simulated trajectory, and that one may benefit from being creative in choosing which discrete-time velocity one applies to certain kinetic measures, such as kinetic energy. Specifically, if one uses SV or VV expressions to approximate a trajectory, it is preferential to apply the half-step LF velocity for kinetic averages, leaving the VV variable $v^n$ as a companion variable for evaluating $r^n$ in the VV algorithm.


\begin{thebibliography}{99}
\bibitem{note_on_typo} This expression for the GJF velocity is a corrected version of the one given in Eq.~(13) of Ref.~\cite{GJF2}.
\end{thebibliography}
\end{document}